# Intra-Family Transformation of The Bi-Te Family via in-situ Chemical Interactions

Zhihao HE [1], Tin Seng Manfred HO [1], Rolf LORTZ [1], and Iam Keong SOU [1,2]*

[1] *Department of Physics, the Hong Kong University of Science and Technology, Hong Kong 999077, China*

[2] *William Mong Institute of Nano Science and Technology, the Hong Kong University of Science and Technology, Hong Kong 999077, China*

* *Corresponding author.*

## Abstract

The Bi-Te binary system, characterized by the homologous series of the $(Bi_2)_m(Bi_2Te_3)_n$, has always attracted research interest for its layered structures and potential in advanced materials applications. Despite $Bi_2Te_3$ has been extensively studied, exploration of other compounds has been constrained by synthesis challenges. This study reports the molecular beam epitaxy (MBE) growth of FeTe on $Bi_2Te_3$, demonstrating that varying growth conditions can turn the $Bi_2Te_3$ layer into different Bi-Te phases and form corresponding FeTe/Bi-Te heterostructures. Our combined analysis using reflection high-energy electron diffraction (RHEED), high-resolution X-ray diffraction (HRXRD), and high-resolution scanning transmission electron microscopy (HR-STEM), indicates that specific growth conditions used for the growth of the FeTe layer can facilitate the extraction of Te from $Bi_2Te_3$, leading to the formation of $Bi_4Te_3$ and $Bi_6Te_3$. Additionally, by lowering the FeTe growth temperature to 230 $^{o}$C, Te extraction from the $Bi_2Te_3$ layer could be avoided, preserving the $Bi_2Te_3$ structure. Notably, all the three FeTe/Bi-Te structures exhibit superconductivity with the FeTe/$Bi_2Te_3$ heterostructure enjoying the highest superconductivity quality. These findings introduce a novel method for realizing $Bi_4Te_3$ and $Bi_6Te_3$ through Te extraction by growing FeTe on $Bi_2Te_3$, driven by the high reactivity between Fe and Te. This approach holds promise for synthesizing other members of the Bi-Te series, expanding the functional potential of these materials.

Keywords: Bi-Te system, Growth of FeTe on $Bi_2Te_3$, Fe reactivity, HRXRD profiling, HR-STEM imaging, Superconducting heterostructure.

## Introduction

$Bi_2Te_3$ has been extensively investigated for several decades as a crucial thermoelectric material [1, 2] and as a topological insulator (TI) [3, 4]. However, other layered Bi-Te compounds have received comparatively less attention [5-7]. The Bi-Te family actually encompasses at least seven distinct crystal structures with the different layered atomic arrangements. The compounds in this family exhibit either rhombohedral or trigonal symmetry, enjoying hexagonal lattices with similar in-plane lattice constants (approximately 0.44 nm) and their lattice constants along c-axis vary from 1 nm to 15 nm [5, 8]. The pioneering work of Stasova and Karpinskii [9] established the presently accepted criterion that all compounds within the Bi-Te family can be described as combinations of $Bi_2$ bilayers (BLs) and $Bi_2Te_3$ quintuple layers (QLs) in the form of $(Bi_2)_m(Bi_2Te_3)_n$. These QLs and BLs are interconnected through van der Waals forces. For instance, the most encountered phase in the Bi-Te system, $Bi_2Te_3$, can be represented as $(Bi_2)_0(Bi_2Te_3)_3$, with a unit cell consisting of 15 atomic layers arranged as QL−QL−QL. It belongs to the rhombohedral space group $R\bar{3}m$ with a hexagonal in-plane lattice [10]. $Bi_4Te_3$, can be expressed as $(Bi_2)_3(Bi_2Te_3)_3$, comprising 21 atomic layers arranged as BL−QL−BL−QL−BL−QL and it also belongs to the rhombohedral space group $R\bar{3}m$ with a hexagonal in-plane lattice [11] Similarly, $Bi_6Te_3$ could be described as $(Bi_2)_2(Bi_2Te_3)_1$, shaped as BL-BL-QL, while its structure belongs to the space group of P-3m1 [12, 13].

$Bi_2Te_3$ has been experimentally demonstrated to be a topological insulator (TI), leading to its extensive study and application in innovative fields such as spintronics and quantum computing [3, 4]. Recent studies have provided evidence supporting that the other compounds in the Bi-Te family also have the potential to exhibit topological insulator behaviours. Yang et al. demonstrated that a single Bi bilayer possesses two-dimensional TI properties, characterized by the presence of nontrivial topological edge states [14]. Based on this finding, they proposed that $Bi_4Te_3$ would also have potential to exhibit TI behaviors. Saito et al. [15] through band structure calculations, predicted that $Bi_4Te_3$ is a bulk zero-band gap semimetal with a Dirac cone at the Γ point. This prediction was subsequently confirmed through optical measurements. Additionally, Eschbach et al. demonstrated that BiTe would exhibit a three-dimensional TI behavior charactered by spin- and angle-resolved photoemission spectroscopy [16]. These recent findings highlight the potential for various Bi-Te compounds to exhibit intriguing topological properties.

Studies of the natural existing minerals of $Bi_6Te_3$, $Bi_7Te_3$ and $Bi_8Te_3$ at least can be dated back to more than 20 years ago [17]. In the realm of experimental synthesis for Bi-Te compounds beyond $Bi_2Te_3$, several efforts have been documented. For instance, $Bi_4Te_3$ nanowires have been successfully synthesized by us via the molecular beam epitaxy (MBE) technique [18]; The realization of $Bi_4Te_3$–ZnTe heterostructure nanowires has been achieved through physical vapor transport (PVT) [19] and BiTe, and $Bi_4Te_3$ thin films have been fabricated utilizing both pulsed laser deposition (PLD) [20] and PVT [21]. Despite these progresses, the experimental fabrication of Bi-Te compounds other than $Bi_2Te_3$ remains limited, particularly in their single crystalline forms. This limitation impedes research into the unique properties of Bi-Te family compounds.

Iron chalcogenides, specifically FeX (X=Se, S, and Te) have attracted significant attention due to their simple chemical composition [22], uncomplicated crystallographic structure [23], and potential for high-temperature superconductivity [24] FeSe and FeS have been demonstrated to display superconductivity [25, 26], while FeTe is reported to be non-superconducting in bulk [27] and thin films [28], as well as under high pressure [29]. It has been reported that long-term exposure to ambient air or annealing in oxygen can induce superconductivity in FeTe. The critical temperatures (Tc) depend on the duration of oxidation or annealing in an oxygen environment [28, 30, 31]. In our previous work, it was demonstrated that superconductivity (SC) emerges at the interfaces of two molecular-beam-epitaxy (MBE)-grown heterostructures composed of topological insulators (TIs) and FeTe ($Bi_2Te_3$/FeTe [32] and $Sb_2Te_3$/FeTe [33]). In both heterostructures, FeTe was used as the bottom layer. The original aim of this study was to cultivate an inverted version of structures, specifically FeTe/$Bi_2Te_3$ or FeTe/$Sb_2Te_3$. This is a crucial step towards achieving structures that contain two or multiple interfaces, such as FeTe/TIs/FeTe, TIs/FeTe/TIs, TIs/FeTe/TIs/FeTe, and so on since they are promising to enjoy enhanced superconductivity due to their quasi-three-dimensional nature. It is generally believed that reducing the growth temperature of the upper layer compared to that of the bottom layer of a heterostructure could decrease the reaction between the two layers. Based on our previous findings, we grew $Bi_2Te_3$/FeTe and $Sb_2Te_3$/FeTe heterostructures at different temperatures. Specifically, $Bi_2Te_3$ was grown at a higher temperature of approximately 240°C while $Sb_2Te_3$ was grown at around 110℃. To avoid the decomposition of the bottom $Sb_2Te_3$ layer, the growth temperature for FeTe in the FeTe/$Sb_2Te_3$ inverted heterostructure should be even lower than 110℃. However, this may result in the formation of amorphous or polycrystalline phases of FeTe. Therefore, we decided to fabricate the FeTe/$Bi_2Te_3$ inverted heterostructure instead. However, prior to this work, a direct growth of FeTe on $Bi_2Te_3$ had not been successfully demonstrated due to the challenges arising from the strong reactivity of Fe with $Bi_2Te_3$, as well as the lattice mismatch between the four-fold symmetry of FeTe and the six-fold symmetry of $Bi_2Te_3$. Regarding the strong interaction between Fe and Te, we have observed two examples previously. The first example is the observation that $Bi_2Te_3$ fluxes interact with Fe islands could induce the formation of FeTe [34]. The second example is our reported study regarding the attempt of the growth of a heavily-Fe-doped $Bi_2Te_3$ thin film, which resulted in a multicomponent product containing not only $Bi_2Te_3$ but also FeTe nanostructures and $FeTe_2$ nanorods [35]. These observations may be attributed to the more negative formation energy of FeTe (-0.270 eV/atom [36]) as compared to that of $Bi_2Te_3$ (-0.259 eV/atom [37]).

Very recently, Yao et al. reported the successful fabrication of FTS (Fe(Te, Se))/$Bi_2Te_3$ heterostructures using the MBE technique [38]. To grow the $Bi_2Te_3$ layer, they initiated the growth with 3 quintuple-layers (QLs) at a relatively low temperature of 240 ℃, which acted as the seed layer. This was followed by the growth of 7 QLs at 300 ℃. The FTS layer was then fabricated by co-evaporating Fe, Te, and Se sources with pre-adjusted fluxes at the same growth temperature of 300 ℃. Based on this report, we aimed to investigate the feasibility of producing FeTe/$Bi_2Te_3$ heterostructures with the FeTe layer grown at high temperatures. We grew FeTe on $Bi_2Te_3$ using similar conditions as Yao's work. Interestingly, we observed that the Bi-Te layer at the bottom of the heterostructure transformed from $Bi_2Te_3$ to $Bi_4Te_3$ or $Bi_6Te_3$ depending on the Fe/Te ratio used during the growth of the top FeTe layer. By lowering the substrate temperature for the growth of the top FeTe layer to 230 ºC, along with appropriate Fe and Te cell temperatures, the FeTe/$Bi_2Te_3$ inverted heterostructure was successfully fabricated. All three FT-BT heterostructures display superconductivity with the FeTe/$Bi_2Te_3$ inverted heterostructure enjoying the highest superconductivity quality, exhibiting superconductivity at approximately 12K and reaching zero resistance at around 7K. These findings demonstrate a potential approach for fabricating various components of the Bi-Te system and shed light on the realization of a TI-FT material system with multiple interfaces, which is promising for achieving higher superconductivity than in its corresponding heterostructure.

# Methods

## Materials Synthesis

The FeTe/Bi-Te (FT-BT) samples and a pure $Bi_2Te_3$ thin film utilised in this study were synthesised in a VG-V80H MBE system. In all FT-BT samples, the $Bi_2Te_3$ layers were grown using a high-purity $Bi_2Te_3$ compound source, while FeTe layers were grown via co-evaporation of high-purity iron lumps and tellurium pieces. All source materials were stored in standard effusion cells. For Sample FT-BT-1, prior to growth, an epi-ready semi-insulating GaAs(111)B substrate was preheated to 580°C to remove the passive oxidation from its surface. Following this, a ZnSe buffer layer was deposited using a ZnSe compound source at a source temperature of 720°C. The subsequent growth of the $Bi_2Te_3$ layer involved a two-step growth: the first step was conducted at a lower growth temperature of 240°C for five minutes, followed by the growth of 295°C for 40 minutes, maintaining the $Bi_2Te_3$ source temperature at 450°C throughout. This two-step growth mode was employed to achieve high structural quality in the as-grown $Bi_2Te_3$ [38]. Subsequently, FeTe was deposited on the $Bi_2Te_3$ at the same temperature of 295 °C for 40 minutes, with Fe source temperature and Te source temperature of 1180 °C and 280 °C, respectively. A pure $Bi_2Te_3$ thin film was fabricated using the same growth conditions established for the $Bi_2Te_3$ layer of the FT-BT-1 heterostructure, however, without the further growth of FeTe.

Sample FT-BT-2 and Sample FT-BT-3 utilised c-plane $\alpha$-$Al_2O_3$(001) (sapphire) wafers as the substrates due to their smoother surface and lower cost as compared to GaAs substrates. Before the growth, the $\alpha$-$Al_2O_3$(001) substrates were preheated at 600°C in the growth chamber for 40 minutes, resulting in clear streaky reflection high-energy electron diffraction (RHEED) patterns. The growth conditions for the $Bi_2Te_3$ layers in both Sample FT-BT-2 and FT-BT-3 were identical. $Bi_2Te_3$ layers were grown on the substrate at 250°C using a similar two-step growth mode as that used in the growth of $Bi_2Te_3$ layer in FT-BT-1 (a lower temperature of 240°C was used for the first 5 minutes, followed by 250°C for 40 minutes, with $Bi_2Te_3$ source temperature at 450°C). The RHEED patterns observed for the as-grown $Bi_2Te_3$ layers on the sapphire wafer are similar to those observed for the $Bi_2Te_3$ layer grown on the GaAs (111)B substrate, indicating that all $Bi_2Te_3$ layers have the same crystalline quality. Following the $Bi_2Te_3$ layer growth, Sample FT-BT-2 underwent the deposition of an FeTe layer at 250°C for 40 minutes, with the Fe and Te source temperatures set to 1150°C and 260°C, respectively. For Sample FT-BT-3, the FeTe layer was grown at a temperature of 230°C for 50 minutes, with the Fe source temperature at 1145°C and the Te source temperature at 255°C. These adjusted temperatures for the Fe and Te sources in these two samples were to prevent the formation of amorphous and polycrystalline FeTe layers.

### Materials Characterization

The crystalline properties of FT-BT and pure $Bi_2Te_3$ samples were monitored during growth using the reflection high-energy electron diffraction (RHEED) system integrated into the MBE system. High-resolution X-ray diffraction (HRXRD) measurements were performed on the as-grown samples using a PANalytical Multipurpose X-ray Diffractometer equipped with Cu Kα1 X-rays (wavelength of 1.54056 Å). To reveal the structure and composition of the FT-BT samples, we conducted cross-sessional high-resolution spherical-aberration-corrected scanning transmission electron microscopy (HR-STEM) imaging using a JEM-ARM200F transmission electron microscope operating at 200 keV. The TEM system is equipped with a probe corrector and a high-angle annular dark-field (HAADF) detector, which allow for detailed analysis of the crystal structure of the samples. All FT-BT samples were cut into long strips using a diamond scriber with a strip size approximately equal to 2 mm × 4 mm. Aluminium wires were bonded onto the surface of the strips to create electrical contacts. The electrical resistances were measured using a standard 4-probe technique in a low temperature cryostat, with measurements taken from 1.4 K to 300 K. The Keithley 6221 AC current source was used in combination with an SR830 lock-in amplifier.

## Results and discussion

The MBE growth of Sample FT-BT-1 was monitored *insitu* using the RHEED technique. During the growth of the $Bi_2Te_3$ layer, its RHEED patterns, as shown in Figure 1(a), displayed a narrower set of streaks and a wider set of streaks when the incident electron beam was along the $[1\bar{1}00]$ and $[11\bar{2}0]$ axes of $Bi_2Te_3$, respectively. The sample rotational angle was defined as $\varphi$=0° and $\varphi$=30° for the two axes, respectively. The spacing ratio for the two sets of streaks (marked in blue and green) is about $1:\sqrt{3}$. The streaky patterns exhibited a 60° rotational symmetry, as expected from the rhombohedral structure of $Bi_2Te_3$ [39]. Figure 1(b) shows the RHEED patterns of the as-grown FeTe layer at rotational angles $\varphi$=0°, $\varphi$=15°, and $\varphi$=30°. These diffraction patterns display a 30° rotational symmetry. The narrower set of streaks and the wider set of streaks occur at 0° and 15° respectively, with a spacing ratio of about 1:$\sqrt{2}$ (marked in blue and red in Figure 1(b)). These observations are similar to those presented in Yao et al.'s report [38]. They found that the as-grown Fe(TeSe) layer consists of three tetragonal lattices twisted by 60°. This is attributed to the six-fold symmetry of the bottom $Bi_2Te_3$ lattice and the four-fold symmetry of the top FeTe or Fe(TeSe) lattice with respect to the growth direction, and their uniaxial lattice matching [38].

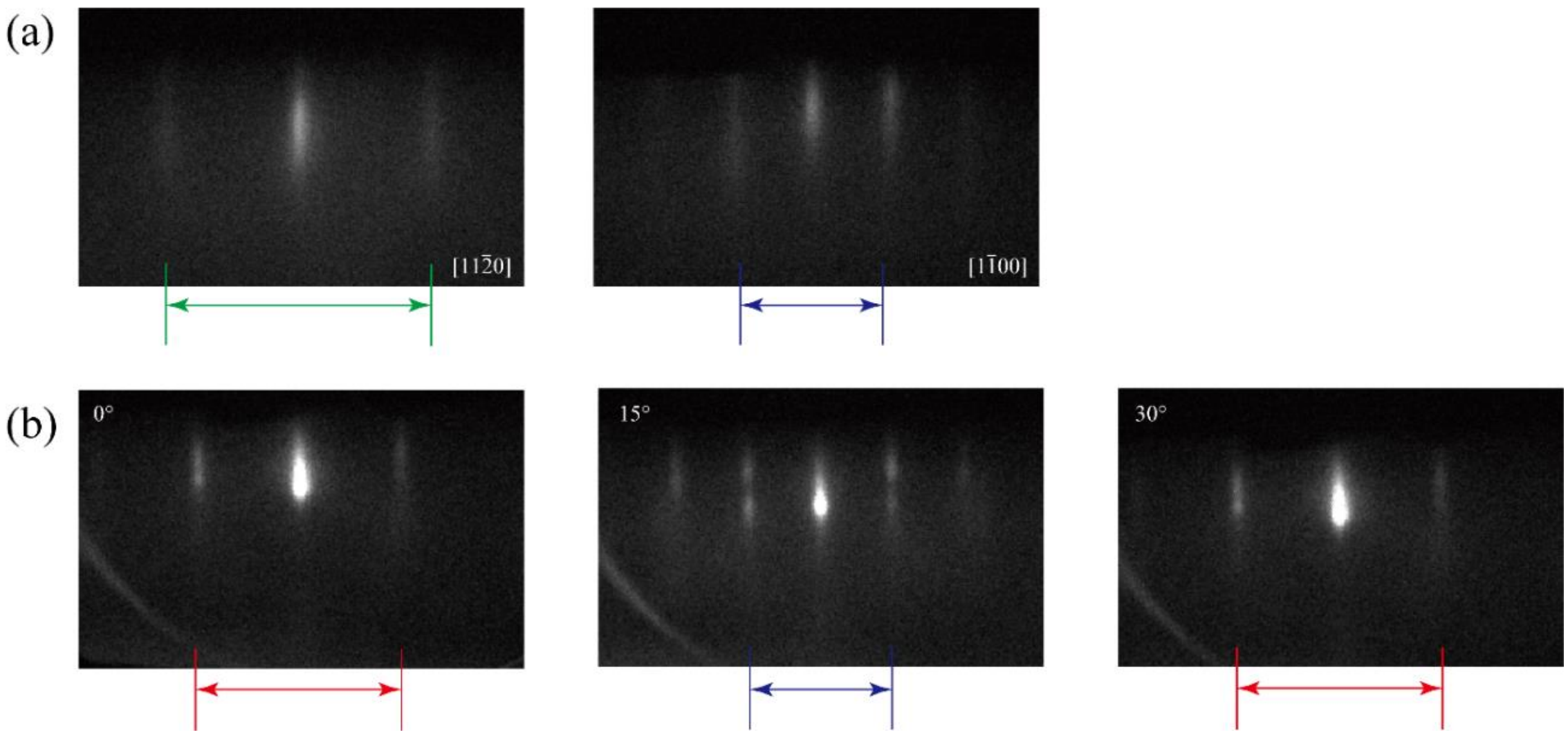

Figure 1. RHEED patterns of (a) $Bi_2Te_3$ layer and (b) as-grown FeTe layer in Sample FT-BT-1, respectively.

In Figure 2(a), lattice structures of FeTe grown on a hexagonal structure with lattice constants similar to that of $Bi_2Te_3$ are displayed. Three tetragonal FeTe lattices with different twisted angles of 0°, 60°, and 120° are named as FT-1, FT-2, and FT-3, respectively, while their unit cell vectors are marked in red, green, and blue. At a rotational angle of 0°, the electron beam is parallel to the [100] direction of FT-1, resulting in a narrower set of streaks in the diffraction pattern. At a rotational angle of 15°, the electron beam is parallel to the [1$\bar{1}$0] direction of FT-2, resulting in a diffraction pattern with a wider set of streaks. At a rotational angle of 30°, the electron beam is aligned parallel to the [0$\bar{1}$0] direction of FT-3, producing the diffraction pattern with a narrower set of streaks. Since the coexistence of FT-1, FT-2, and FT-3 lattices, the overall RHEED patterns of the FeTe layer grown on $Bi_2Te_3$ display a 12-fold symmetry. We have modified the description of the occurrence of this 12-fold symmetry, based on Yao et al.'s explanations [38]. Assuming that the bottom layer of the as-grown heterostructure is $Bi_2Te_3$, its in-plane lattice constant is approximately 4.4 Å, while the in-plane lattice constant of FeTe is about 3.78 Å [32], thus they follow a lattice spacing ratio of $2:\sqrt{3}\approx 8:7$. Due to this ratio, the bottom Te layer of each of the three FeTe lattices described above could almost perfectly align with the top Te layer of the $Bi_2Te_3$ lattice. Figure 2(b) demonstrates a good alignment between the FT-1 lattice and the $Bi_2Te_3$ lattice. Each of the four sides of the bottom Te layer of FeTe with 8 x 8 units (marked in red) aligns well with the underlying $Bi_2Te_3$ (marked in blue) lattice, with 8 lattice constants of FeTe matching well with 7 lattice constants of $Bi_2Te_3$. The same close match also occurs for the other two FeTe lattices (FT-2 and FT-3) with $Bi_2Te_3$, as shown in Figure S1 in the supplementary information.

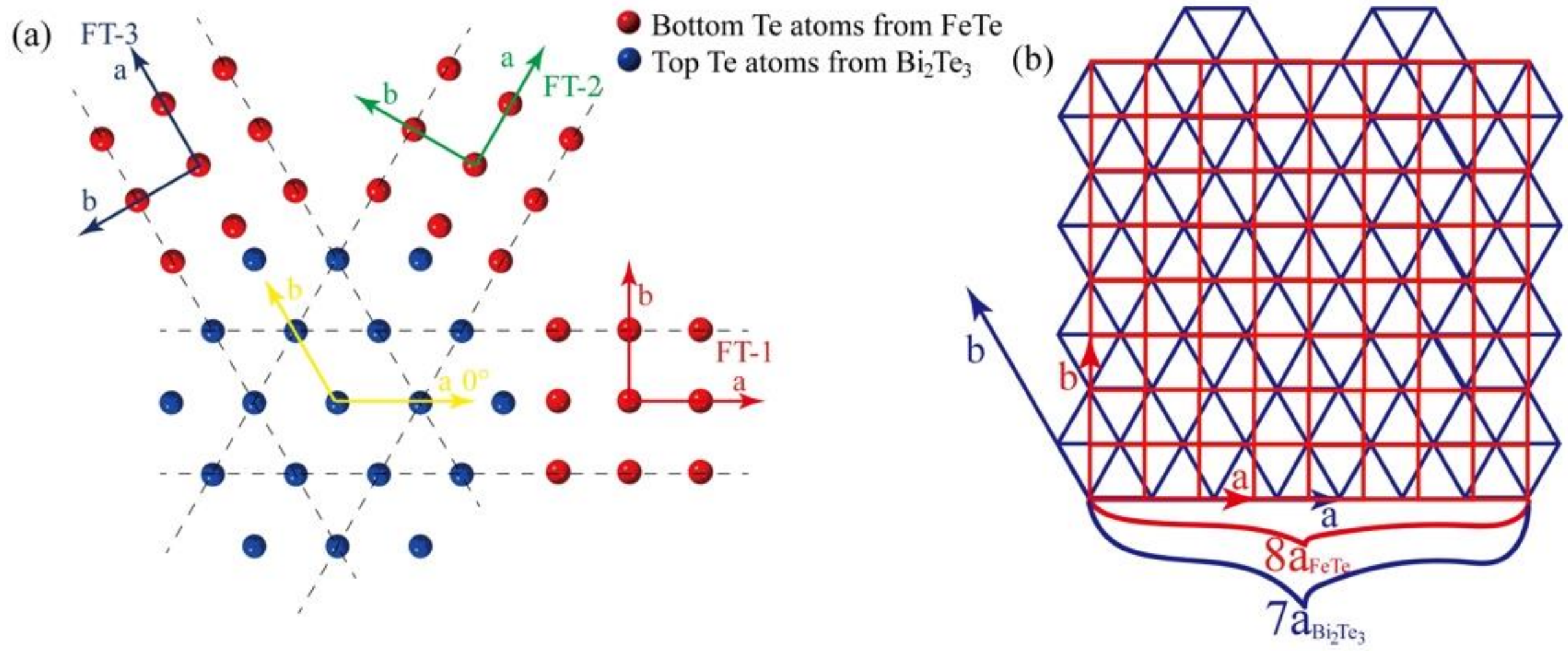

Figure 2. Schematic drawings of an $FeTe/Bi_2Te_3$ inverted heterostructure. (a) The three lattice structures of FeTe with different in-plane orientations grown on $Bi_2Te_3$. (b) The FT-1 lattice lying on the $Bi_2Te_3$ lattice with the relation of 8 lattice constants of FeTe matching well with 7 lattice constants of $Bi_2Te_3$.

Figure 3(a) shows the HRXRD profile of Sample FT-BT-1. Figure 3(a) shows that Sample FT-BT-1 has characteristic diffraction peaks from FeTe (0 0 m) and peaks that may be attributed to either $Bi_2Te_3$ (0 0 n) or $Bi_4Te_3$ (0 0 n), in addition to the peaks from the substrate. It is important to note that the 3-index notation, which is commonly used in HRXRD profiles, is equivalent to the 4-index notation used in the RHEED patterns presented in Figure 1. Figure 3(b) and (c) show the zoom-in HRXRD profiles of Sample FT-BT-1. The selected ranges cover the peak of FeTe (001) (from 13° to 16°) and the peaks of $Bi_2Te_3$ (0 0 15) and $Bi_4Te_3$ (0 0 21) (from 44° to 47°). The FeTe (001) peak of Sample FT-BT-1 is located at approximately 14.14°, corresponding to a lattice constant of about 6.26 Å, which is very close to that of bulk FeTe (6.28 Å, extracted from PDFs #29-0729). Figure 3(c) shows that Sample FT-BT-1 has a peak at approximately 45.275° from the Bi-Te layer, which is significantly different from the peak position of 44.576° of the (0 0 15) peak of bulk $Bi_2Te_3$. Upon further investigation of the peak positions of other diffraction peaks of the Bi-Te layer of Sample FT-BT-1, it was found that they are all close to the 2θ values of the diffraction peaks of $Bi_4Te_3$. For instance, the peak at 45.275° is in proximity to the (0 0 21) peak of bulk $Bi_4Te_3$, which has a standard 2θ value of 45.449° (from PDFs #33-0216). The observed 2θ value of this diffraction peak is slightly lower than that of bulk $Bi_4Te_3$, most likely due to the thin-film strain effect on our sample.

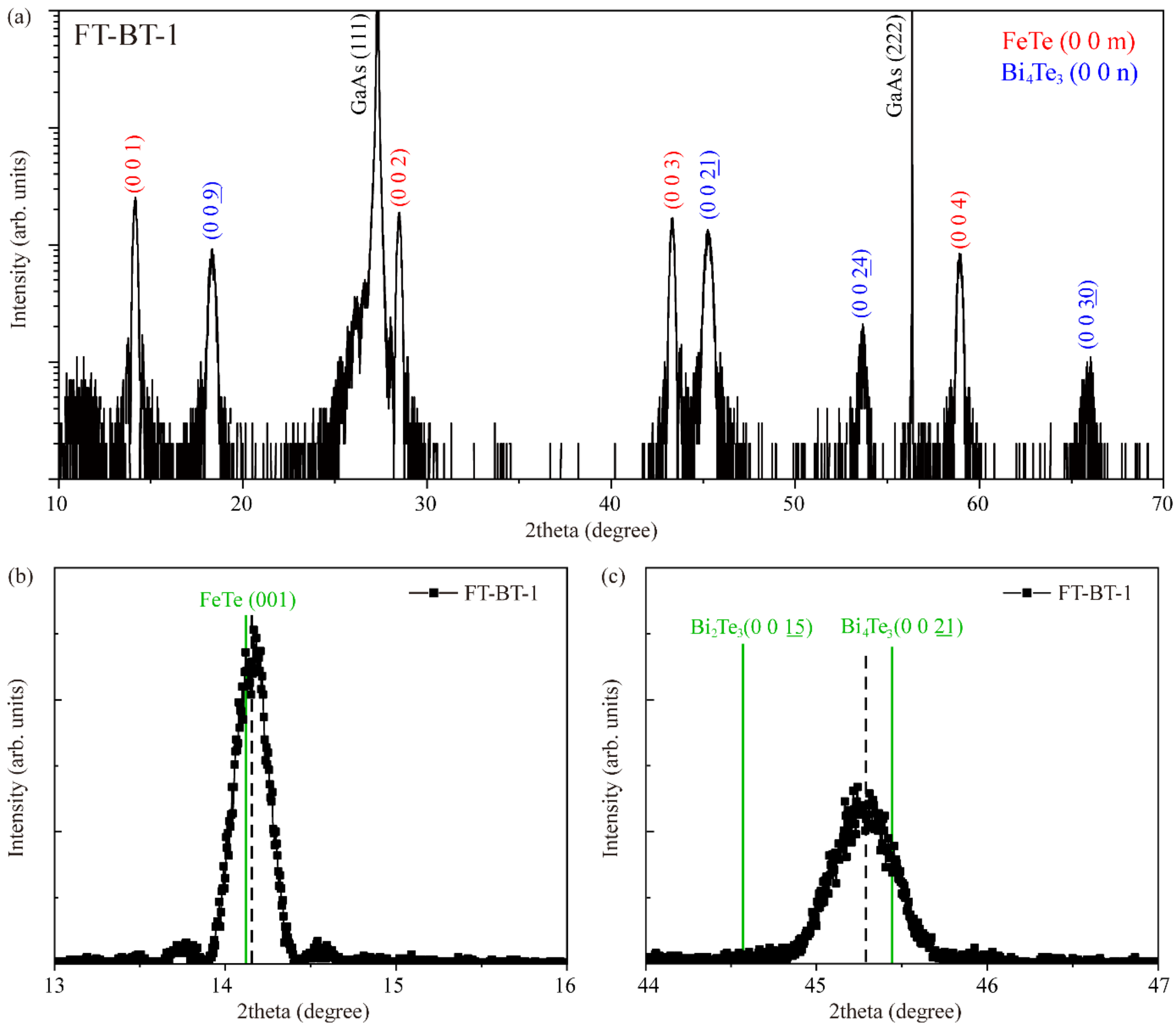

Figure 3. (a) High-resolution X-ray diffraction (HRXRD) profile of Sample FT-BT-1 from 10° to 70°. Zoom-in HRXRD profiles of Sample FT-BT-1 within the ranges (b) from 13° to 16°, and (c) from 44° to 47°, where green solid lines indicate the positions of peaks from FeTe (0 0 1), $Bi_2Te_3$ (0 0 15), and $Bi_4Te_3$ (0 0 21), respectively.

To obtain detailed information on the structure and composition of the Bi-Te layer in Sample FT-BT-1, cross-sectional HR-STEM imaging was performed, which provides precise images and allows for analysis of the microstructure with atomic-scale resolution. Figure 4(a) shows a HR-STEM image of Sample FT-BT-1, revealing that the thicknesses of the Bi-Te layer and FeTe layer are approximately 18 nm and 14 nm, respectively. Figure 4(b) presents HR-STEM images captured at two distinct regions with a larger magnification. The top image shows the region near the interface, while the bottom image corresponds to the region within the Bi-Te layer, where the zone axis is aligned along the $[11\bar{2}0]$ direction of $Bi_2Te_3$ or $Bi_4Te_3$. The images show short red lines with two neighbouring lines confining a septuple unit comprising one bilayer of Bi and one quintuple layer of $Bi_2Te_3$, and the eight units defined by the short red lines are labelled as T1-T4 in the top image and B1-B4 in the bottom image. All eight septuple layers match the expected atomic arrangement of $Bi_4Te_3$, indicating that Sample FT-BT-1's Bi-Te layer is indeed $Bi_4Te_3$.

Notably, at the top image of Figure 4(b), we constructed an angle α, which is the acute angle between the horizontal line along the $[11\bar{2}0]$ direction and a line through seven closely packed atoms, measured to be about 65°. This is larger than the value of about 42° measured in theoretical models. The deviation is attributed to the fact that the $Bi_4Te_3$ layer in the FT-BT-1 sample was formed through Te extraction from $Bi_2Te_3$ during the growth of FeTe, rather than through a strict thermal equilibrium growth, details of which are addressed later in this paper. However, it is important to note that this deviation does not affect the structural integrity of $Bi_4Te_3$, where three septuple layers still constitute a unit cell, which is evidenced in the top image of Figure 4(b), in which one can see that the atomic arrangements in septuple layers T1 and T4 are identical.

In the bottom image of Figure 4(b), the periodic arrangements from B1 to B4 appear to be similar, which is different from what we have addressed for the top image of Figure 4(b). We believe that this inconsistence may be caused by the deformation of the thin specimen used to take the bottom image of Figure 4(b), which is evidenced by comparing the atomic arrangement with a straight dash line marked in blue in this image. As can be seen, this line marking the direction through seven closely packed atoms in the lower part of this image cannot align with the atoms in the upper part, indicating that the lattice appears to be twisted toward the left at the upper part.

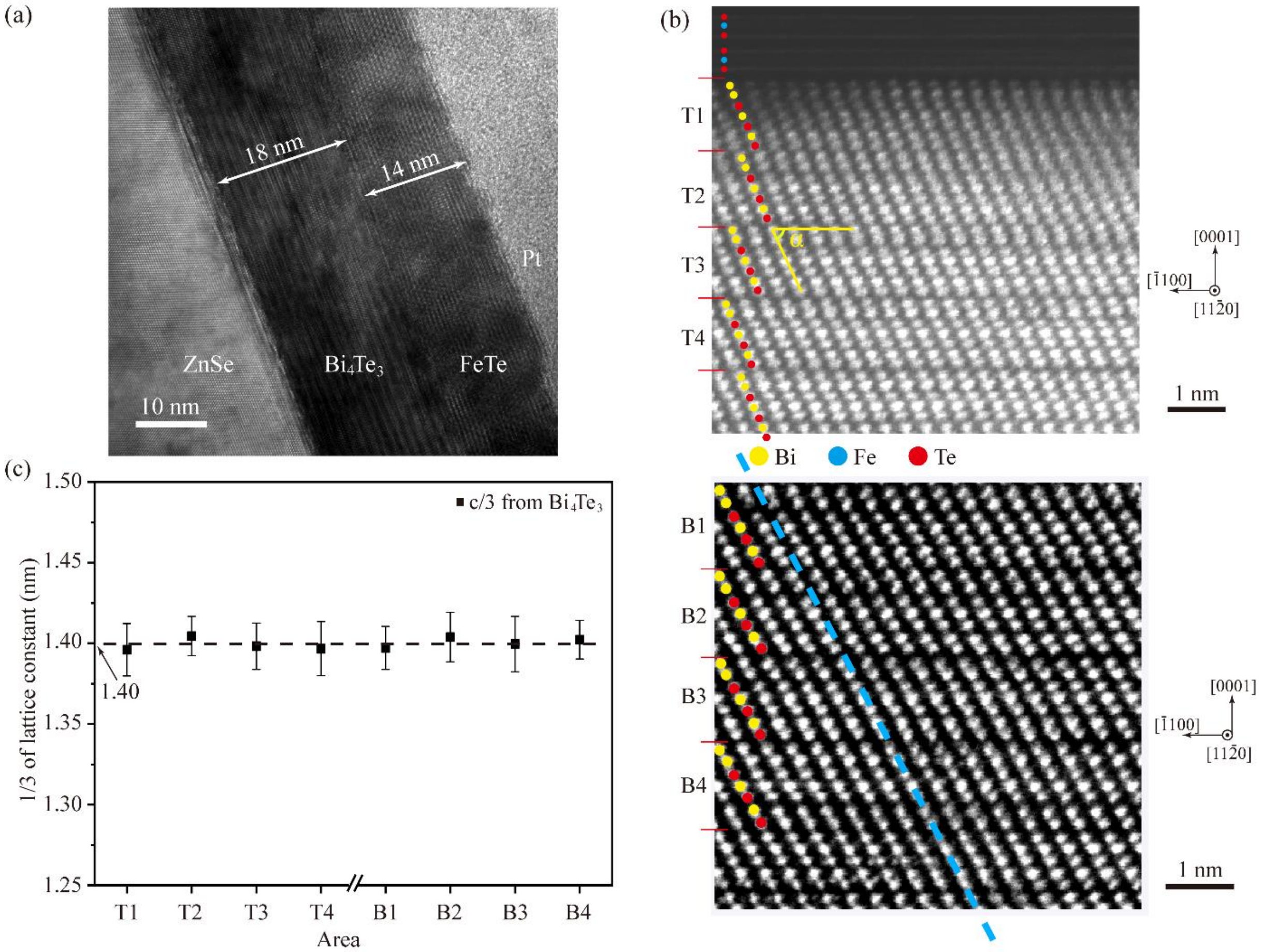

Figure 4. Structural analysis of Sample FT-BT-1. (a) Its cross-sectional HR-STEM image. (b) Higher-magnification cross-sectional HR-STEM images of the heterostructure captured in two regions: the interface (top) and the Bi-Te layer (bottom), with the zone axis aligned along the [11$\bar{2}$0] direction of $Bi_4Te_3$. Short red lines indicate the 1/3 spacing of the unit cells. The areas with sufficient resolution and uniform atomic arrangements are labelled as T1-T4 in the top image and B1-B4 in the bottom image. (c) Variation of the lattice parameters of $Bi_4Te_3$, extracted from the HR-STEM images in (b), where the error bars specify the standard deviations of the measurements. The dash line indicates that the fitted mean c/3 value of $Bi_4Te_3$ is about 1.40 nm.

Figure 4(c) presents the variation of the lattice parameters derived from the HR-STEM images shown in Figure 4(b), with error bars indicating the standard deviations of the measurements. The dashed line indicates the fitted c/3 value for $Bi_4Te_3$, which is approximately 1.40 nm. This value reflects that the lattice constant along the c-axis of the $Bi_4Te_3$ layer in Sample FT-BT-1 is approximately 42.0 Å, closely matching the calculated value of 42.027 Å obtained from the HRXRD profile displayed in Figure 3. This provides strong evidence that the HRXRD peaks from the Bi-Te layer of Sample FT-BT-1 are attributed to a $Bi_4Te_3$ layer and further confirms that this Bi-Te layer is indeed $Bi_4Te_3$.

In order to understand the formation mechanism of the $Bi_4Te_3$ layer in Sample FT-BT-1, we first aimed to eliminate the possibility that the as-grown Bi-Te layer is inherently composed of $Bi_4Te_3$ prior to the growth of the top FeTe. To achieve this, we epitaxially grew an individual $Bi_2Te_3$ thin film under the same growth conditions used for the fabrication of the BT layer in Sample FT-BT-1. Figure 5 displays the HRXRD profile of this $Bi_2Te_3$ thin film, where the characteristic layer peaks match well with the peaks of bulk $Bi_2Te_3$, indicating that the $Bi_4Te_3$ layer in Sample FT-BT-1 was transformed from the $Bi_2Te_3$ layer during the growth of the top FeTe layer due to the high reactivity between Fe and Te. It is believed that during FeTe growth, some Te atoms from the $Bi_2Te_3$ bottom layer diffuse to the growth front and react with arriving Fe atoms to form part of the top FeTe layer. As mentioned earlier, this is based on the fact that the FeTe enjoys a more negative formation energy than that of $Bi_2Te_3$. In more detail, it could be considered that during the growth of the top FeTe layer, three Te atomic layers from every other $Bi_2Te_3$ unit of the bottom $Bi_2Te_3$ layer were extracted to the top, leaving behind two Bi atomic layers. These two Bi atomic layers then combine with the neighbouring $Bi_2Te_3$ unit to form the Bi-Bi-$Bi_2Te_3$ unit of $Bi_4Te_3$.

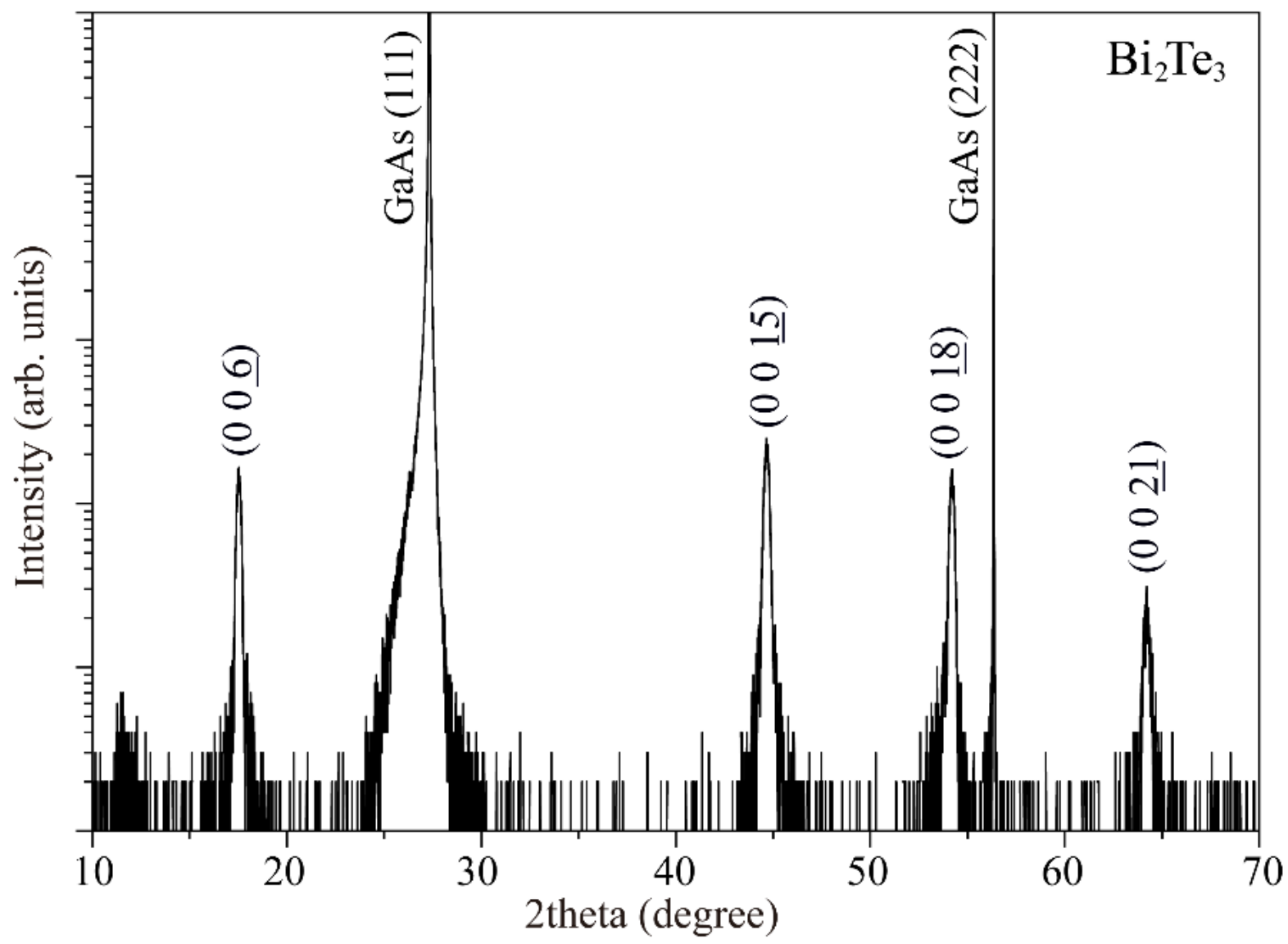

Figure 5. HRXRD profile of a $Bi_2Te_3$ film grown under the same growth conditions used for growing the $Bi_2Te_3$ layer in Sample FT-BT-1.

After performing structural characterizations on Sample FT-BT-1 and determining that the composition of the BT layer in FT-BT-1 is $Bi_4Te_3$, we fabricated Sample FT-BT-2 by reducing the substrate temperature and lowering the temperatures of the Fe and Te sources for the growth of the top FeTe layer. Specifically, the top FeTe layer was fabricated at about 250°C, with the Fe source temperature at 1150°C and the Te source temperature at 260°C. Comparing to the source temperatures used for the growth of FT-BT-1, the reduction in Te flux (64.7%) was more substantial than the reduction in Fe flux (38.8%) as derived from the relation between the vapour pressure and cell temperature for the elemental Te and Fe [40]. It is believed that this factor overwhelms the effect of reducing the substrate temperature as the MBE growth rate of FeTe is well known to be determined by the Fe flux [30]. Consequently, this adjustment of the source temperatures used for the growth of Sample FT-BT-2 is expected to enhance the extraction of Te from the $Bi_2Te_3$ layer and the as-grown FeTe may also suffer from having more Fe interstitials as well.

During the growth of Sample FT-BT-2, the similar evolutions of RHEED patterns was observed as previously described for Sample FT-BT-1. Figure 6 displays the HRXRD profiles of FT-BT-2 from 10° to 70°, including the zoom-in profiles within the ranges from 13° to 16°, and from 44° to 47°. The zoom-in XRD profiles shown in Figure 6(b) reveals that the FeTe (001) peak of Sample FT-BT-2 is located at about 14.09°, corresponding to a lattice parameter of about 6.28 Å, which is almost the same as that of bulk FeTe (6.28 Å, extracted from PDFs #29-0729). Regarding the peaks originating from the BT layer, as shown in Figure 8(c), a peak in the zoom-in range at 44.65° is observed, which is significantly far away from the $Bi_4Te_3$ (0 0 21) peak at 45.449° but close to the position of the bulk $Bi_2Te_3$ (0015) peak at 44.576°. However, HR-STEM images and data analysis of FT-BT-2 as shown in Figure 7 indicate that the main composition of BT layer in Sample FT-BT-2 is neither $Bi_2Te_3$ nor $Bi_4Te_3$, but $Bi_6Te_3$.

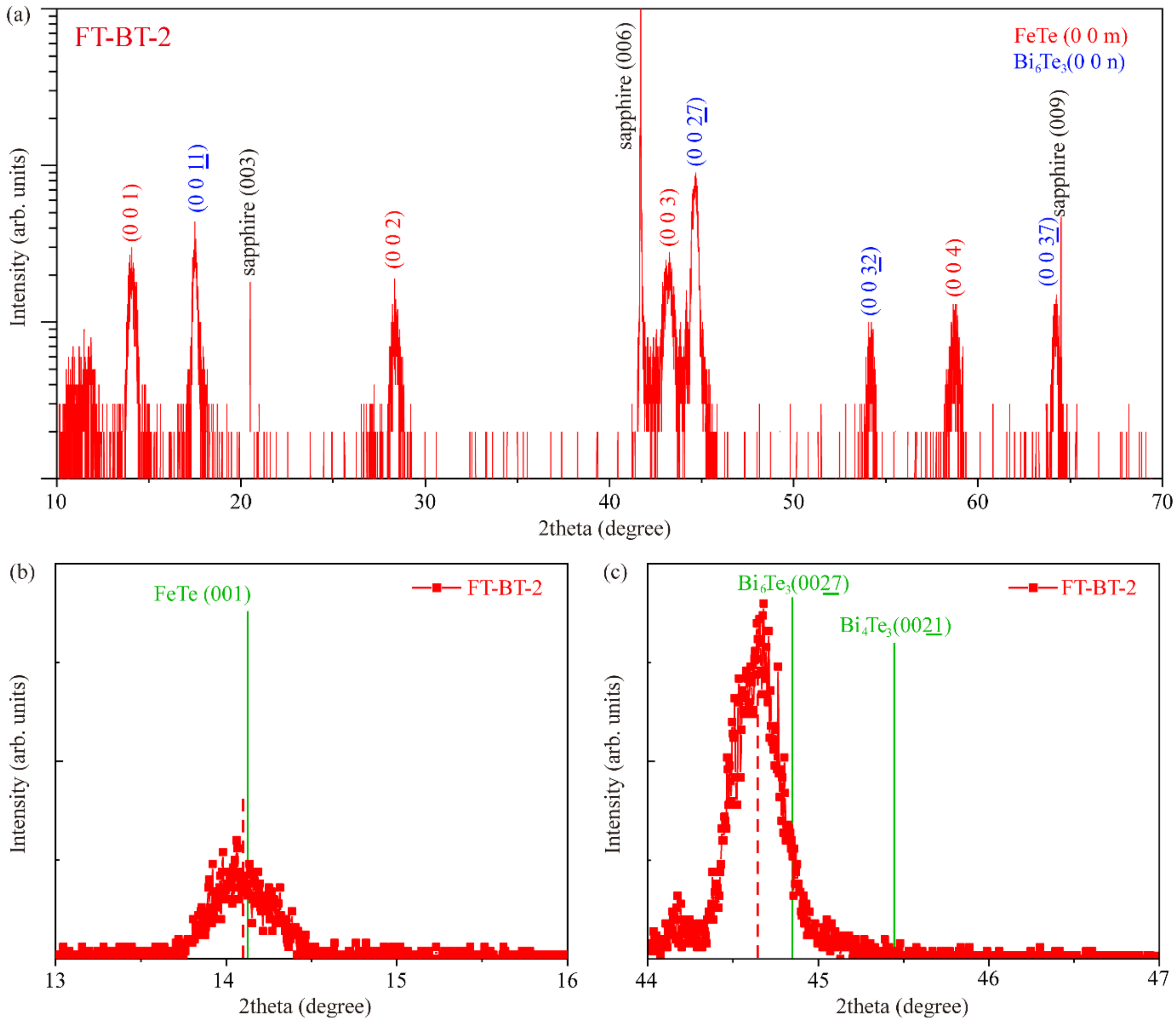

Figure 6.(a) High-resolution X-ray diffraction (HRXRD) profiles of FT-BT-2 from 10° to 70°. Zoom-in HRXRD profiles of FT-BT-2 within the ranges (b) from 13° to 16°, and (c) from 44° to 47°, where green solid lines indicate the positions of standard peaks of FeTe (001), $Bi_6Te_3$ (0 0 $\underline{27}$), and $Bi_4Te_3$ (0 0 $\underline{21}$).

Figure 7(a) shows a HR-STEM image of Sample FT-BT-2, revealing that the thicknesses of the Bi-Te layer and FeTe layer are approximately 28 nm and 15 nm, respectively. Figure 7(b) shows HR-STEM images that reveal the presence of nonuple layers consisting of $Bi_6Te_3$, another distinct member in the Bi-Te system. It can be regarded as having quadruple atomic layers of bismuth between each $Bi_2Te_3$ slab or two bismuth bi-layers positioned at each side of $Bi_2Te_3$ to form a lattice arrangement shaped as $(Bi_2)_2(Bi_2Te_3)_1$. In theoretical calculations, $Bi_6Te_3$ enjoys a similar in-plane lattice constant with $Bi_4Te_3$ and $Bi_2Te_3$ (around 4.4 Å) but is categorized under the space group P-3m1, distinct from the $R\bar{3}m$ space group of $Bi_4Te_3$ and $Bi_2Te_3$ [12, 13]. $Bi_6Te_3$'s nonuple layer itself constitutes a unit cell, in contrast to $Bi_4Te_3$ and $Bi_2Te_3$, where a unit cell is made up of three neighboring septuple or quintuple layers respectively. Despite the theoretical clarity, experimental fabrications of $Bi_6Te_3$ have not been extensively documented, leaving some characteristics of this material less understood and they may arise due to the use of a specific growth method. As observed in the bottom image of Figure 7(b), there is a slight shift of each nonuple $Bi_6Te_3$ layer along the $[\bar{1}\bar{1}20]$ direction from the nonuple layers of B3 to B1. Interestingly, it can be seen that the lattice arrangement of the B0 nonuple layer repeats the lattice arrangement of B3 nonuple layer~~,~~ suggesting that the unit cell of the $Bi_6Te_3$ layer in Sample FT-BT-2 comprises three neighbouring nonuple layers. This difference between theoretical calculations and our experimental observations might be attributed to the fact that the formation of $Bi_6Te_3$ in this heterostructure was induced by a high extraction rate of Te atoms from the $Bi_2Te_3$ layer, which definitely was undergone a non-thermal-equilibrium process that could offer a possibility to generate a product deviating from the expectation of theoretical calculations. It is worth mentioning that in the top image of Figure 7(b), one can find quite a lot of Fe interstitials as marked by the red rectangles in the FeTe layer of Sample FT-BT-2, which echoes what has been predicted earlier based on the higher Fe/Te ratio used for the growth the top FeTe of this heterostructure.

The same methods used in Sample FT-BT-1 are employed to measure the lattice parameters of the $Bi_6Te_3$ layer in Sample FT-BT-2. As shown in Figure 7(c), extracted spacing of 1/3 lattice period along the c-axis of $Bi_6Te_3$ is determined to be approximately 1.81 nm. This value agrees well with the previously calculated value of one $Bi_6Te_3$ nonuple layer (18.09 Å) [13]. Consequently, the peaks in the HRXRD profile of Sample FT-BT-2 could be assigned to $(0\ 0\ \underline{11})$, $(0\ 0\ \underline{27})$, $(0\ 0\ \underline{32})$, and $(0\ 0\ \underline{37})$ peaks of $Bi_6Te_3$, respectively. Notably, the peak corresponding to $Bi_6Te_3$ $(0\ 0\ \underline{27})$ located at 44.65° in the HRXRD profile suggests that its lattice constant along the c-axis of one unit cell is about 54.66 Å, being close to the value of 54.3 Å as estimated from its HR-STEM images.

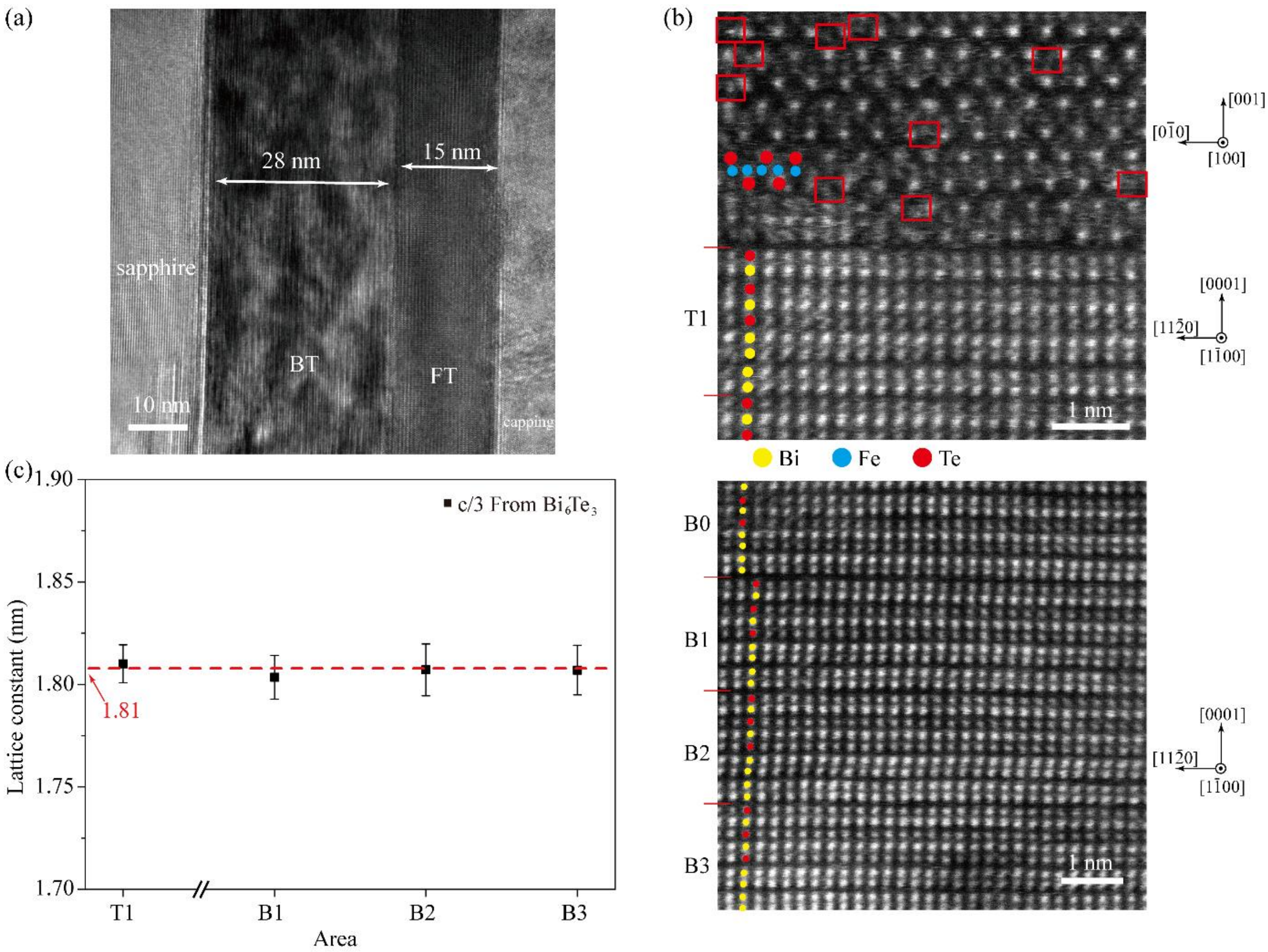

Figure 7. Structural analysis of Sample FT-BT-2. (a) Its cross-sectional HR-STEM image. (b) Higher-magnification cross-sectional HR-STEM images of the heterostructure captured in two regions: the interface (top) and the Bi-Te layer (bottom), with the zone axis aligned along the [1$\bar{1}$00] direction of $Bi_6Te_3$. Short red lines indicate the 1/3 spacing of the unit cells. The areas with sufficient resolution and uniform atomic arrangements are labelled as T1 in the top image and B1-B4 in the bottom image. (c) Variation of the lattice parameters of $Bi_6Te_3$, extracted from the HR-STEM images in (b), where the error bars specify the standard deviations of the measurements. The dash line indicates that the fitted mean c/3 value of $Bi_6Te_3$ is about 1.81 nm.

Regarding the formation mechanism of the $Bi_6Te_3$ layer of Sample FT-BT-2, it is recalled that the top FeTe layer of Sample FT-BT-2 was grown using a higher Fe/Te ratio as compared with that of Sample FT-BT-1. Thus we believe that, during the growth of the top FeTe layer of Sample FT-BT-2, more Te atoms of the bottom $Bi_2Te_3$ layer were expected to get extracted to the top as compared with the growth of Sample FT-BT-1, leading to the formation of $Bi_6Te_3$ via the combination of four Bi layers and a unit of $Bi_2Te_3$, which is a more Bi-richer BT layer than the $Bi_4Te_3$ layer in Sample FT-BT-1.

After determining the structures and compositions of the BT layers in Sample FT-BT-1 and Sample FT-BT-2, we proceeded to fabricate Sample FT-BT-3 by further reducing the substrate temperature and adjusted the temperatures of the Fe and Te sources. These adjustments were based on the hypothesis that the extraction of Te from $Bi_2Te_3$ results from the high reactivity between Fe and Te during the growth of the top FeTe layer. By lowering the substrate temperature, we aimed to ascertain whether the use of a lower substrate temperature for the growth of FeTe could mitigate the reaction and preserve the integrity of the structure of the bottom $Bi_2Te_3$ layer.

During the growth of Sample FT-BT-3, a similar evolution of RHEED patterns was observed as those of Sample FT-BT-1 and Sample FT-BT-2. Figure 8 shows the HRXRD profile of Sample FT-BT-3 from 10° to 70°, including the zoom-in profiles. The HRXRD profiles in Figure 8 (b) show that the FeTe (001) peak of FT-BT-3 is located at approximately 14.22°, corresponding to a lattice parameter of approximately 6.22 Å, which is close to the value of bulk FeTe (6.28 Å, extracted from PDFs #29-0729). As shown in Figure 6(c), the peak located at about 44.7° matches well with the (0 0 15) peak of bulk $Bi_2Te_3$ (at 44.576°, extracted from PDFs #15-0863). This value corresponds to the lattice constant along c-axis about 30.37 Å, which is also close to the standard value of bulk $Bi_2Te_3$ (30.48 Å, extracted from PDFs #15-0863).

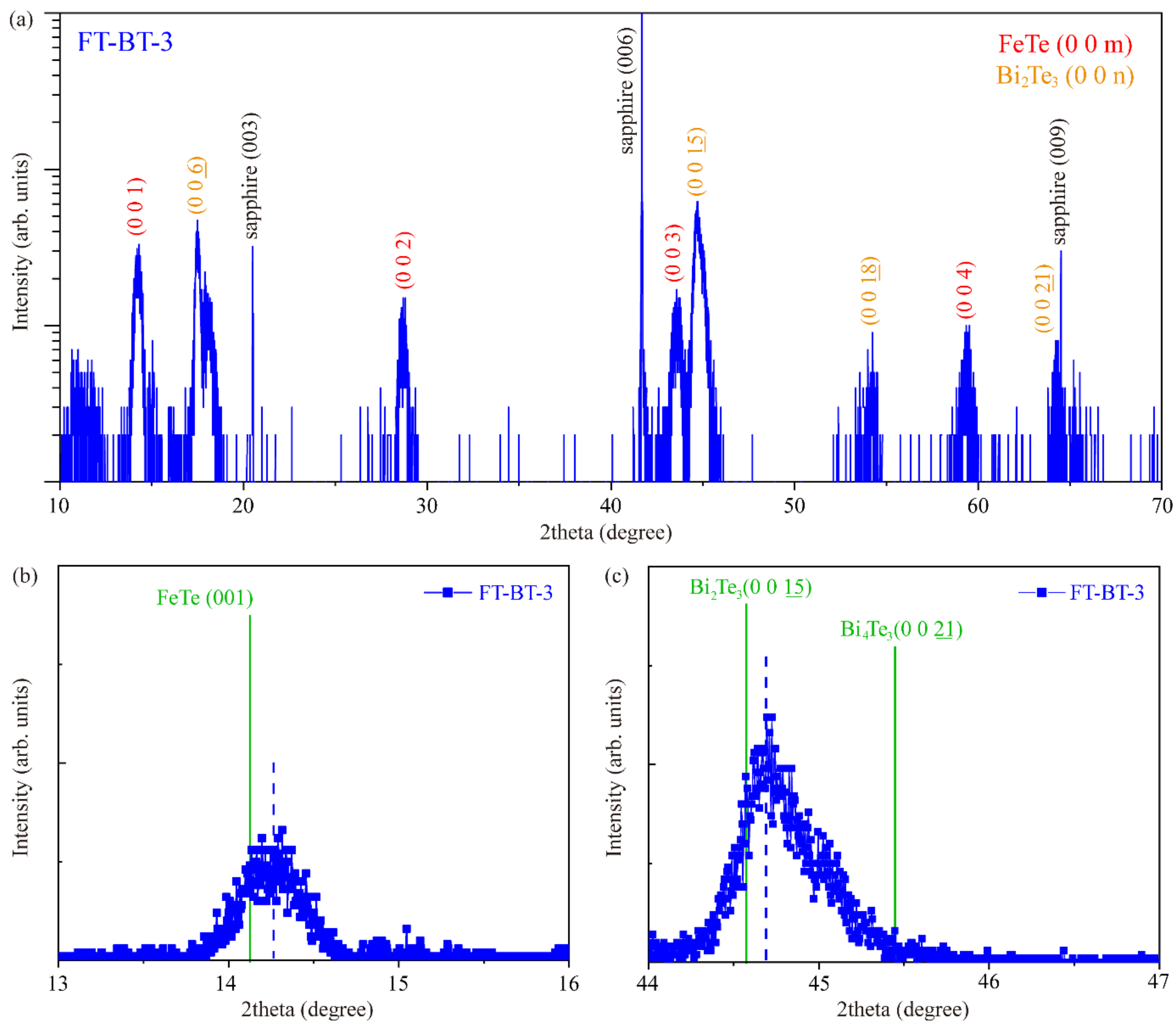

Figure 8. (a) High-resolution X-ray diffraction (HRXRD) profile of Sample FT-BT-3 from 10° to 70°. Zoom-in HRXRD profiles of Sample FT-BT-3 within the ranges (b) from 13° to 16°, and (c) from 44° to 47°, where green solid lines indicate positions of standard peaks of FeTe (0 0 1), $Bi_2Te_3$ (0 0 15), and $Bi_4Te_3$ (0 0 21).

Figure 9 displays the HR-STEM images and the data analysis for Sample FT-BT-3. Figure 9(a) shows the cross-sectional HR-STEM image of this sample, revealing the single-crystalline lattices of each layer of the FeTe/$Bi_2Te_3$ heterostructure and the sapphire substrate. Figure 9 (b) presents the HR-STEM images of a region in the BT layer, highlighting its atomic structure and the van der Waals gaps (marked by the red lines), which correspond well with the expected features of a $Bi_2Te_3$ layer. We analysed the composition of the $Bi_2Te_3$ layer in Sample FT-BT-3 using the same method as for Sample FT-BT-1 and Sample FT-BT-2. Figure 9(c) shows the fitted c/3 value extracted from the HR-STEM image of the $Bi_2Te_3$ layer of Sample FT-BT-3 (Fig. 9 (b)) to be about 0.995 nm, indicating that the lattice constant along the c-axis of a unit cell of the $Bi_2Te_3$ layer is of about 29.85 Å, which is also close to the standard value of bulk $Bi_2Te_3$ (30.48 Å).

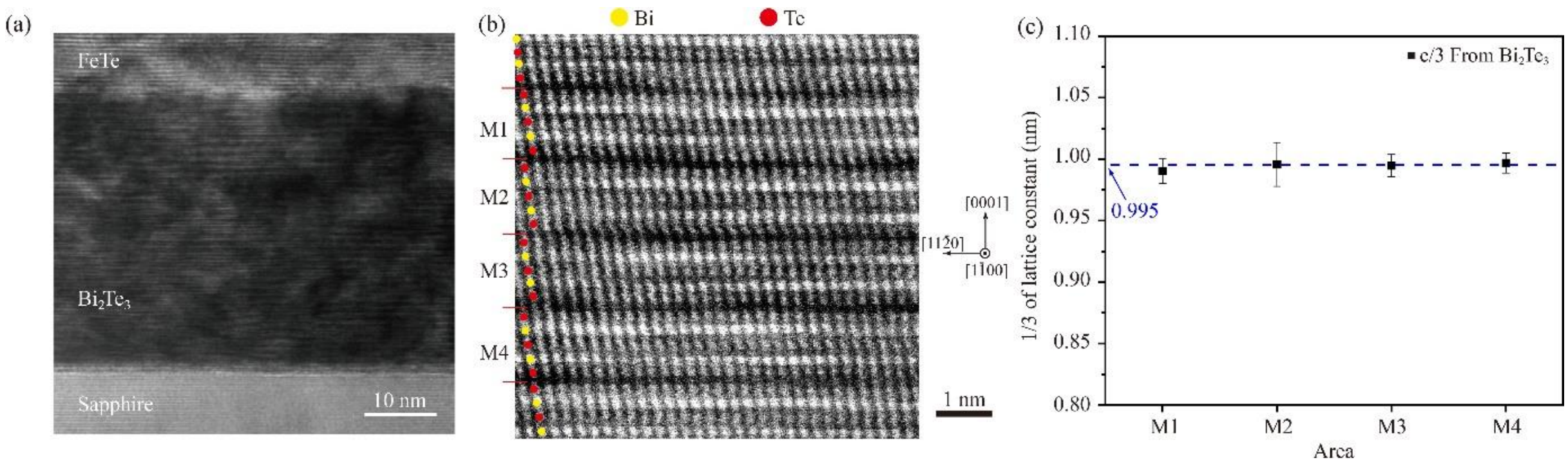

Figure 9. Structural analysis of Sample FT-BT-3. (a) Its cross-sectional HR-STEM image. (b) Higher-magnification cross-sectional HR-STEM images of the $Bi_2Te_3$ layer. Short red lines indicate the 1/3 spacing of the unit cells. The areas with sufficient resolution and uniform atomic arrangements are labelled as M1-M4. (c) Variation of the lattice parameters of $Bi_2Te_3$, extracted from the HR-STEM image in (b), where the error bars specify the standard deviations of the measurements. The dash line indicates the fitted mean c/3 value of $Bi_2Te_3$ is about 0.995 nm.

The supplementary information includes a HR-STEM image of Sample FT-BT-3 at a rare region (Figure S2). In this region, a thin FeTe sheet is visible between the sapphire substrate and the $Bi_2Te_3$ layer. However, it is important to note that most of the interface right above the sapphire substrate does not show such a thin sheet. The reason why Sample FT-BT-3 allows the formation of FeTe sheets at some rare regions above the surface of the substrate is likely attributed to the lower substrate temperature used for the growth of the top FeTe layer. During the growth of the top FeTe layer of this sample, dissociation of the bottom $Bi_2Te_3$ layer was absent. Consequently, no significant outgoing Te flux helped to block the incoming Fe atoms entering the bottom $Bi_2Te_3$ layer due to diffusion. Most of the diffusing Fe atoms accumulate at the bottom of the $Bi_2Te_3$ layer and dissociate a portion of the $Bi_2Te_3$ layer to form rare FeTe sheets. It is important to note that a desired inverted heterostructure of $Bi_2Te_3$/FeTe has been achieved under the growth conditions used for fabricating Sample FT-BT-3, except that some rare FeTe sheets exist above the substrate surface. The successful realization of the FeTe/$Bi_2Te_3$ heterostructure as demonstrated in this study puts one step closer for fabricating multilayer structures that feature multiple interfaces between $Bi_2Te_3$ and FeTe, which are expected to display enhanced superconductivity due to its quasic three-dimensional nature.

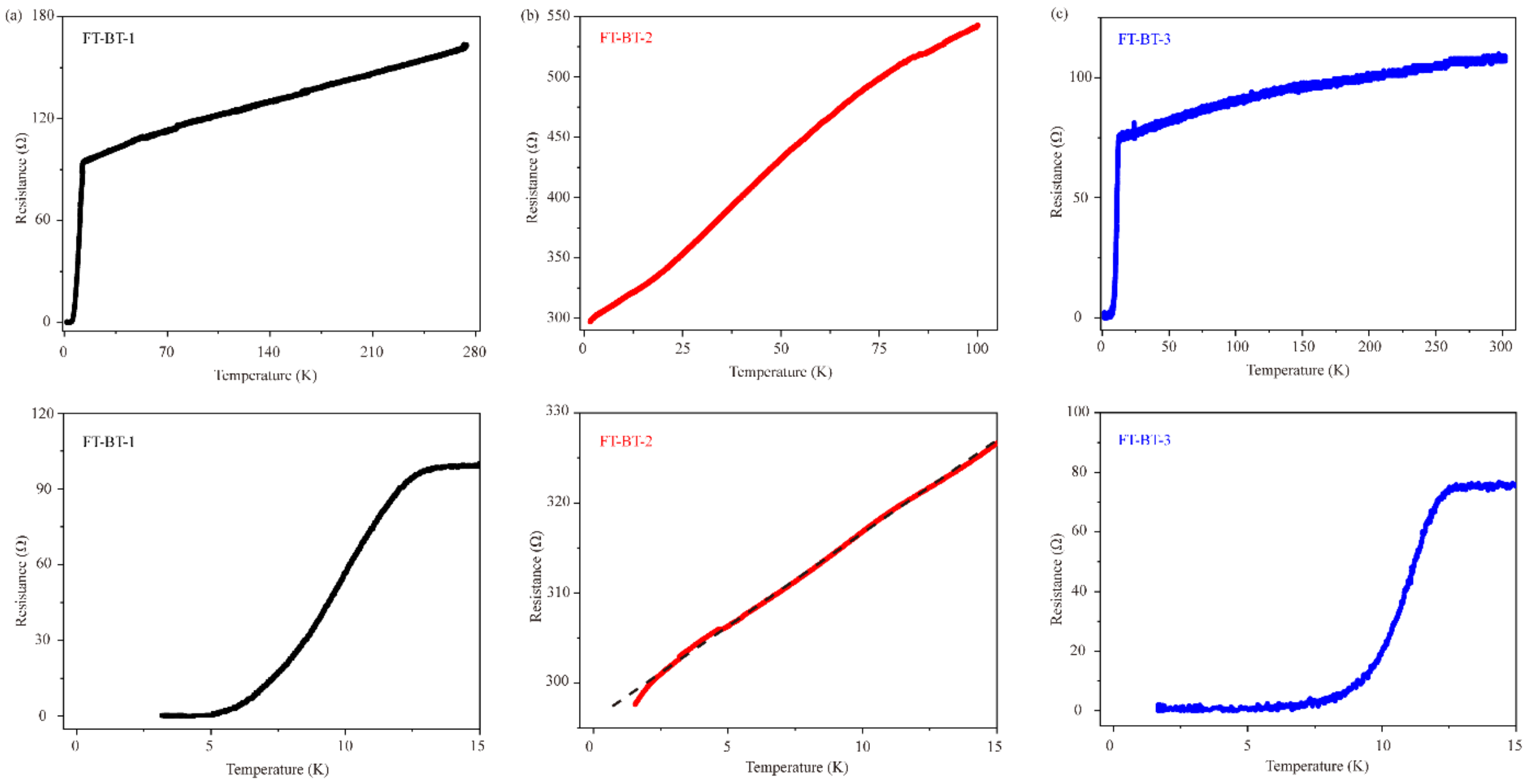

Figure 10. Temperature-dependent resistance curves of (a) FT-BT-1, (b) FT-BT-2, and (c) FT-BT-3, respectively.

To determine the superconductivity of the three FT-BT samples, temperature-dependent electrical resistance measurements (R-T) were conducted using a commercial Quantum Design SQUID magnetometer and the results are displayed in Figure 10, As can be seen in this figure, the R-T curve of Sample FT-BT-1 shows a notable resistance drop occurs around 12K, and the resistance reaches zero at approximately 5K. For Sample FT-BT-2, only a weak superconductivity signal appears at 2.3K as reflected by a small fall of its resistance, which is believed to be correlated to the high concentration of Fe interstitials in its top FeTe layer as addressed earlier. This in fact echoes our previous finding that Fe interstitials in FeTe could adversely affect the quality of the superconductivity in $Sb_2Te_3$/FeTe heterostructures [33]. Regarding Sample FT-BT-3, the superconducting transition begins at 12 K, with resistance reaching zero at about 7 K, displaying a sharper drop as compared to that of Sample FT-BT-1. It is believed that the observed superconductivity in FT-BT-3 mainly originates from the interface between the bottom $Bi_2Te_3$ layer and the top FeTe layer, while the superconductivity arising from the interface between the FeTe sheets and the $Bi_2Te_3$ layer is a minor contribution. This is because if the latter was the dominating contribution, the resistance of this heterostructure would not reach zero, since the rare FeTe sheets are spatially separated from each other as addressed earlier.

## Conclusion

In conclusion, we conducted a series of FeTe growth on $Bi_2Te_3$ by the MBE technique through varying the substrate temperatures alongside modifications to the source cell temperatures, resulting in three typical FT-BT samples grown under different growth conditions. The structures of the three samples were analysed via RHEED observation, HRXRD profiling, and HR-STEM imaging. Our analysis reveals that specific growth conditions of the top FeTe layer could lead to the extraction of Te atoms from the $Bi_2Te_3$ layer, facilitating the transformation of $Bi_2Te_3$ into $Bi_4Te_3$, and with a higher Fe/Te ratio, even into $Bi_6Te_3$. Moreover, a lower growth temperature of FeTe could avoid Te extraction from $Bi_2Te_3$ and form a FeTe/$Bi_2Te_3$ heterostructure although some FeTe sheets were found to form in rare regions above the substrate surface. It was found that all three FT-BT samples display superconductivity with the FeTe/$Bi_2Te_3$ sample enjoying the highest superconductivity quality. These studies reveal a new approach to realizing the $Bi_4Te_3$ and $Bi_6Te_3$ components of the Bi-Te family by extracting Te atoms from a $Bi_2Te_3$ thin film based on the strong reactivity between Fe and Te. It is expected that this approach may also be applied to generate other components of this material system.

## Acknowledgements

We gratefully acknowledge the use of the facilities in the Materials Characterization and Preparation Facility (MCPF) at the Hong Kong University of Science and Technology. The work described here was substantially supported by funding from the Research Grants Council of the Hong Kong Special Administrative Region, China, under Grant Numbers 16308020, C6025-19G.